%% file: main.tex
\documentclass[aps,prl,amsmath,amssymb,latexsym,array,enumerate,letter,twocolumn,superscriptaddress]{revtex4-1}
\usepackage{amssymb}
\usepackage{amsmath}
\usepackage{epsfig}
\usepackage{hyperref}
\usepackage{breakurl}
\usepackage{mathtools}
\usepackage{extpfeil}
\usepackage{comment}
\usepackage{multirow}
\usepackage{graphicx,float,color}
\usepackage{url}
\usepackage{ulem} 
\usepackage{newfile}
\usepackage{orcidlink}
\makeatletter

\def\mnras{Mon. Not. Roy. Astron. Soc.}

\def\apjl{Astrophys. J. Lett.}
\def\aap{Astron. Astrophys.}

\def\prd{Phys. Rev. D}

\begin{document}

\title{Equation of State of Decompressed Quark Matter, and Observational Signatures of Quark-Star Mergers}

\author{Zhiqiang Miao\orcidlink{0000-0003-1197-3329}}
\email{zq\_miao@sjtu.edu.cn}
\affiliation{Tsung-Dao Lee Institute, Shanghai Jiao Tong University, Shanghai, 201210, China}

\author{Zhenyu Zhu\orcidlink{0000-0001-9189-860X}}
\affiliation{Tsung-Dao Lee Institute, Shanghai Jiao Tong University, Shanghai, 201210, China}
\affiliation{Center for Computational Relativity and Gravitation, Rochester Institute of Technology, Rochester, NY 14623, USA}

\author{Dong Lai\orcidlink{0000-0002-1934-6250}}
\email{donglai@sjtu.edu.cn}
\affiliation{Tsung-Dao Lee Institute, Shanghai Jiao Tong University, Shanghai, 201210, China}
\affiliation{Department of Astronomy, Cornell Center for Astrophysics and Planetary Science, Cornell University, Ithaca, NY 14853, USA}

\date{\today}

\begin{abstract} 
Quark stars are challenging to confirm or exclude observationally
because they can have similar masses and radii as neutron stars. 
By performing the first calculation of the non-equilibrium equation of state of decompressed quark matter at finite temperature, we determine the properties of the ejecta from binary quark-star or quark star-black hole mergers. 
We account for all relevant physical processes during the ejecta evolution, including quark nugget evaporation and cooling, and weak interactions. We find that these merger ejecta can differ significantly from those in neutron star mergers, depending on the binding energy of quark matter. 
For relatively high binding energies, quark star mergers are unlikely to produce r-process elements and kilonova signals. 
We propose that future observations of binary mergers and kilonovae could impose stringent constraints on the binding energy of quark matter and the existence of quark stars.

\end{abstract}

\maketitle 

{\it Introduction:} Quark stars (QSs) are hypothetical astrophysical objects, motivated by the suggestion that deconfined quark matter may have lower energy than nuclear matter at zero external pressure. Despite decades of speculation~\citep{1971PhRvD...4.1601B,1984PhRvD..30..272W}, it remains difficult to confirm or rule out the existence of QSs~\citep{2005PrPNP..54..193W,2014PhRvD..89d3014D,2016PhRvD..94h3010L,2018PhRvL.120v2001H,2021PhRvL.126p2702B,2022A&A...660A..62T}. 
For typical stellar masses, the mass-radius curves of QSs closely resemble those of neutron stars (NSs), making it challenging to distinguish between them observationally.
Even with recent precise measurements of radii ~\citep{2019ApJ...887L..21R,2019ApJ...887L..24M,2021ApJ...918L..27R,2021ApJ...918L..28M,2024ApJ...961...62V,2024ApJ...974..294S,2024ApJ...971L..20C,2024ApJ...974..295D} or tidal deformabilities~\citep{2019PhRvX...9a1001A}, QSs still fit within the observational constraints~\citep{2018PhRvD..97h3015Z,2021MNRAS.506.5916L,2022MNRAS.515.5071M,2022PhRvD.106h3007C,2022PhRvD.105l3004Y,2020ApJ...905....9T,2021ApJ...917L..22M}. 

The kilonova event AT2017gfo~\citep{2017Natur.551...64A,2017ApJ...848L..17C,2017Sci...358.1556C,2017Sci...358.1570D,2017Sci...358.1565E,2017Natur.551...71T,2017ApJ...848L..27T,2017ApJ...848L..24V,2017ApJ...850L...1L}, associated with the gravitational-wave source GW170817~\citep{2017PhRvL.119p1101A}, is widely believed to originate from a binary neutron star (BNS) merger.
This raises the question: can a binary quark star (BQS) merger produce a similar kilonova?
Kilonovae are powered by the radioactive decay of heavy element isotopes formed via r-process nucleosynthesis in a neutron-rich environment~\citep{1998ApJ...507L..59L,2010MNRAS.406.2650M}.
Therefore, the key question is whether a BQS merger can produce a comparably neutron-rich environment.
Several studies suggest that during a BQS merger, the evaporation of quark matter into nucleons could be very efficient, such that most of the ejected matter is neutron rich and contributes to a kilonova~\citep{2019ApJ...881..122D,2022PhRvD.106j3032B}.
However, these calculations, based on a formalism originally developed for quark nugget evaporation in the early universe~\citep{1985PhRvD..32.1273A}, did not account for the saturation of evaporated nucleons. 
Saturation means that the nuggets and the evaporated nucleons have reached equilibrium, with an equilibrium nucleon density $n^{\rm eq}$. 
For a quark nugget with baryon number $A$, the evaporation rate is $|dA/dt|_{\rm evap}=n^{\rm eq}\langle\sigma v\rangle$, where $\sigma\simeq\pi A^{2/3}\,{\rm fm^2}$ is the nucleon-nugget collision cross section, and $v\sim \sqrt{T/m_n}$ is the thermal velocity of the nucleons. The brackets $\langle\cdot\rangle$ denote a thermal average over the Maxwell-Boltzmann velocity distribution.
When the mean baryon number density is $n_B$, the saturation timescale is of order
\begin{equation}
\begin{split}
    &\tau_{\rm sat} = \frac{n^{\rm eq} A/n_B}{|dA/dt|_{\rm evap}} = \frac{A}{n_B\langle\sigma v\rangle} \\
    &\sim 1.7\times 10^{-13}\,{\rm s} \left(\frac{0.1\,{\rm fm^{-3}}}{n_B}\right)\left(\frac{10\,{\rm MeV}}{T}\right)^{1/2}\left(\frac{A}{10^{30}}\right)^{1/3}\,.
\end{split}
\end{equation}
For the decompressed ejecta following BQS mergers, $n_B\sim 10^{-3}\,{\rm fm^{-3}},\,T\sim 10\,{\rm MeV},\,A\sim10^{30}$~\citep{2022PhRvD.106j3032B}, leading to $\tau_{\rm sat}\sim 10^{-11}\,{\rm s}$, much shorter than the ejecta expansion time $\tau_{\rm exp}\gtrsim 10^{-3}\,{\rm s}$. 
Consequently, nugget evaporation in BQS mergers can quickly reach saturation, thereby suppressing further evaporation and allowing more quark nuggets to survive.

If a significant number of quark nuggets survive, we must reassess the equation of state (EoS) of decompressed quark matter, which is critical for understanding the outcomes of BQS mergers.
While a few numerical simulations of BQS mergers exist~\citep{2009PhRvL.103a1101B,2010PhRvD..81b4012B,2021PhRvD.104h3004Z,2022PhRvD.106j3030Z,2024arXiv240711143G}, they have primarily focused on the bulk quark matter EoS, where the density exceeds nuclear density, but have paid little attention to the sub-nuclear density regime, which is relevant to the ejecta.
In the simulations, the ejecta EoS is typically described by incorporating an ideal-fluid thermal component, i.e., $P_{\rm th} =(\Gamma_{\rm th}-1)u_{\rm th}$, where $P_{\rm th}$ and $u_{\rm th}$ are the thermal pressure and internal energy density. 
The adiabatic index $\Gamma_{\rm th}$ is chosen to be $7/4$ in Ref.~\citep{2021PhRvD.104h3004Z} and $4/3$ in Ref.~\citep{2022PhRvD.106j3030Z}. 
These choices may not be entirely appropriate, as the thermal pressure contribution in the ejecta depends on the composition of the gas. Therefore, $\Gamma_{\rm th}$ will evolve with both density and temperature, depending on the nugget evaporation process. This evolution needs to be addressed if we aim to develop a robust ejecta evolution model.

Also relevant are NS-black hole (BH) mergers, as several possible NS-BH gravitational-wave events have been reported~\citep{2020ApJ...896L..44A,2021ApJ...915L...5A,2024ApJ...970L..34A}, although their electromagnetic counterparts have not yet been observed~\citep{2019ApJ...880L...4H,2019ApJ...881L...7G,2021NatAs...5...46A}. 
However, as more such events are detected in the future, it is interesting to ask whether there are any differences in the observational signatures between NS-BH and QS-BH mergers.

In this Letter, we address the aforementioned questions by modeling the evolution of the merger ejecta, considering relevant processes like nugget evaporation, nugget cooling, and weak reactions.
For the first time, we compute the non-equilibrium EoS of decompressed quark matter, thus determining the properties of the ejecta from BQS or QS-BH mergers.
We find that depending on the binding energy of quark matter relative to nuclear matter, such mergers can be quite different from NS mergers in terms of producing r-process elements and kilonova.

{\it Equilibrium EoS of decompressed quark matter:} The ejected matter after the BQS merger, referred to as quark nugget gas (QNG), can be divided into two phases: the nugget phase and the nucleon gas phase. Surface tension limits the initial nugget size from becoming too small during turbulent fragmentation, with the estimated baryon number $A$ exceeding $10^{26}$~\citep{2022PhRvD.106j3032B}. In this Letter, we assume $A$ is sufficiently large, so that the specific value of $A$ does not affect our results.

Shortly after the merger, the ejecta is subjected to high temperatures ($\sim 10\,\rm{MeV}$) and high densities ($\sim 10^{12}\,{\rm g/cm^3}$). In this stage, the main physical processes, including nucleon emission and absorption, as well as weak reactions, i.e.,
\begin{align}
    &(A) \leftrightarrow (A-1) +n,\ (A) \leftrightarrow (A-1) +p,\label{eq: A-n}\\
    &n+e^+ \leftrightarrow p+\bar\nu_e,\ p+e^- \leftrightarrow n+\nu_e,\ n \leftrightarrow p+e^-+\bar\nu_e, \label{eq: n to p}
\end{align}
occur rapidly compared to the ejecta expansion.
Thus, the quark nuggets and nucleons in the gas reach an equilibrium state, determined by the balances of chemical potentials and temperatures, and charge neutrality:
\begin{align}
    \mu_p^{(\rm {Q})} +\mu_e^{(\rm {Q})}=\mu_n^{(\rm {Q})} = \mu_n^{(\rm G)}=\mu_p^{(\rm G)} +\mu_e^{(\rm G)}\,,\label{eq: chemical eq}\\
    T^{\rm ({Q})} = T^{\rm (G)}\,,\\
    n_p^{\rm (G)} = n_{e^-}^{\rm (G)}-n_{e^+}^{\rm (G)} = \frac{[\mu_e^{\rm (G)}]^3}{3\pi^2}+\frac{\mu_e^{\rm (G)}[T^{\rm (G)}]^2}{3}\,,\label{eq: charge neutral}
\end{align}
where superscripts ``(G)'' and ``(Q)'' refer to gas and quark nugget phases, respectively.
For simplicity, we will omit the ``(G)'' superscript when referring to the gas phase and let $T_s = T^{\rm ({Q})}$ as the nugget internal temperature. In Eq.~(\ref{eq: charge neutral}) we have used $\mu_{e^+}=-\mu_e$. For the nugget phase, we express the nucleon chemical potential as $\mu_n^{\rm ({Q})} = 3\mu_q$ and $\mu_p^{\rm ({Q})} = 3\mu_q-\mu_e^{\rm ({Q})}$, with $\mu_q$ the average quark chemical potential inside the nugget. 
Eq.~(\ref{eq: chemical eq}) already implies the beta equilibrium outside the nuggets. For non-relativistic and non-degenerate nucleons in the gas, 
\begin{equation}\label{eq: chemical potential}
    \mu_{n,p}-m_{n,p} = T\ln \left[ \left(\frac{2\pi}{m_{n,p}T}\right)^{3/2}\frac{n_{n,p}}{2}\right],
\end{equation}
with $m_{n,p}$ the masses of neutron and proton.
An important quantity is the binding energy of the nugget, $\Delta E =m_n-\mu_n^{\rm ({Q})}$. Given $\Delta E$, solving Eqs.~(\ref{eq: chemical eq}--\ref{eq: chemical potential}) determines the equilibrium state. 

We can compute the EoS of the QNG relating the density, pressure and internal energy density:
\begin{align}
    &\rho = Am_nn_A+m_nn_n+m_pn_p,\\
    &P_{\rm th}=P_n+P_p+P_{e^+e^-}+P_\gamma,\label{eq: pressure}\\
    &u_{\rm th}=u_A +u_n+u_p+u_{e^+e^-}+u_\gamma,\label{eq: e density}
\end{align}
where $n_A$ is the nugget number density and
\begin{align}
    &u_A = \frac{3\pi^2T^2}{2\mu_q}An_A\,,\\
    &P_n = \frac{2}{3} u_n = n_nT\,,\\
    &P_p = \frac{2}{3} u_p = n_pT\,,\\
    &P_{e^+e^-}  =\frac13u_{e^+e^-}
     =\frac{7\pi^2T^4}{180}\left[1+\frac{30}{7\pi^2}\left(\frac{\mu_e}{T}\right)^2+\frac{15}{7\pi^4}\left(\frac{\mu_e}{T}\right)^4\right]\,,\\
    &P_\gamma = \frac13 u_\gamma = \frac{\pi^2T^4}{45}\,.
\end{align}
We neglect the contribution to the pressure from nuggets since $P_A = n_AT \ll P_{\rm th}$ for sufficiently large $A$. 

If the temperature is too high, quark nuggets would completely evaporate into nucleons, at which point the QNG would be replaced by ordinary nucleon gas (NG). This transition occurs when
\begin{equation}\label{eq:critical line}
    \rho = m_nn_n^{\rm eq}+m_pn_p^{\rm eq}\,,
\end{equation}
where $n_n^{\rm eq}$, $n_p^{\rm eq}$ are the equilibrium densities, as discussed above.

{\it Non-equilibrium EoS:} As the ejecta expands and cools, 
the relevant reaction rates in QNG decrease and eventually become slower than the expansion rate, causing the system out of thermal and chemical equilibrium. 
We now describe the thermal and chemical evolution in this regime. 
We no longer assume the nugget temperature $T_s$ is equal to the environment (gas) temperature $T$. 
Instead, $T_s$ is determined by the nugget cooling process, as described by the energy equation:
\begin{equation}
    \frac{d U}{dt} = L_\nu + L_n+L_p,\label{eq: internal energy}
\end{equation}
where $U = 3\pi^2AT_s^2/(2\mu_q)$ is the total internal energy of the nugget, with $\mu_q = (m_n-\Delta E)/3$, and $L_\nu$ and $L_{n,p}$ represents the neutrino cooling rate and nucleon cooling rate, respectively. Following Ref.~\citep{1985PhRvD..32.1273A}, we write these rates as 
\begin{align}
    L_\nu &= 4\pi R_s^2 \left(\frac{7\pi^2}{160}\right)\left[T^4p(R_s, T)-T_s^4p(R_s, T_s)\right]\,,\\
    L_{n,p} &= -\frac{dN_{n,p}}{dt} (\Delta E+ 2T)\,,
\end{align}
where $p(R_s,T)=\frac43R_sG_{\rm F}^{2}\mu_q^2T^3$ and $R_s$ is the nugget size, $N_n=n_n/n_A$ and $N_p=n_p/n_A$ are the numbers of free neutrons and protons per nugget volume, respectively. {$G_{\rm F}$ is the Fermi coupling constant.}

The net nucleon evaporation rate, accounting for both nucleon emission and absorption, are given by 
\begin{align}
    \left[\frac{dN_n}{dt}\right]_{A\leftrightarrow (A-1) + n} &= n_n^{\rm eq}\langle \sigma_n v_n\rangle_{T_s}-n_n\langle \sigma_n v_n\rangle_{T},\label{eq: dNn_eva}\\
    \left[\frac{dN_p}{dt}\right]_{A\leftrightarrow (A-1) + p} &= n_p^{\rm eq}\langle \sigma_pv_p\rangle_{T_s}-n_p\langle \sigma_pv_p\rangle_{T},\label{eq: dNp_eva}
\end{align}
Here $\sigma_{n,p}$ are the nucleon absorption cross-sections and $v_{n,p}$ are the corresponding thermal velocities. 
For neutrons, we adopt the geometric cross-section of the nugget, i.e., $\sigma_n =\pi R_s^2${, since the nugget size ($\sim10^8\,{\rm fm}$) is much larger than the strong interaction scale}. 
However, protons must overcome a Coulomb barrier to enter the nugget~\citep{1986ApJ...310..261A}, making the cross-section energy-dependent, i.e., $\sigma_p = P(E)\sigma_n$, where $P(E)$ is a penetration factor.
The height of the Coulomb barrier at the nugget surface, $E_c$, is roughly a few $\rm MeV$.
Averaging over $E$ yields (see Supplemental Material~\citep{SM})
\begin{equation}
    \langle \sigma_pv_p\rangle_{T}\approx  \sigma_n \sqrt{\frac{8T}{\pi m_p}} \int_\eta^\infty \frac{4\sqrt{1-\eta/\xi}}{(1+\sqrt{1-\eta/\xi})^2}\xi e^{-\xi}d\xi
\end{equation}
where $\xi= E/T$ and $\eta = E_c/T$.

The conversion between neutrons and protons outside the nuggets [cf. Eq.(\ref{eq: n to p})] gives
\begin{equation}
    \left[\frac{dN_n}{dt}\right]_{n\leftrightarrow p}=-\left[\frac{dN_p}{dt}\right]_{n\leftrightarrow p}  = -N_n\lambda_{n
    \to p} +N_p\lambda_{p \to n}\,.\label{eq: np conversion}
\end{equation}
{The corresponding conversion rates are given by~\citep{1983bhwd.book.....S}
\begin{align}
    &\lambda_{n\to p} =\frac{G_{\rm F}^2C_V^2(1+3a^2)}{2\pi^3}\int dE_\nu  E_\nu^2\nonumber\\
    & \times \int dE_eE_e\left(E_e^2-m_e^2\right)^{1/2}f_{e^+} \delta\left(E_\nu-Q-E_e\right) \,, \\
    &\lambda_{p\to n} =\frac{G_{\rm F}^2C_V^2(1+3a^2)}{2\pi^3}\int dE_\nu  E_\nu^2\nonumber\\
    & \times \int dE_eE_e\left(E_e^2-m_e^2\right)^{1/2}f_{e^-} \delta\left(E_\nu+Q-E_e\right) \,,
\end{align}
where $Q=m_n-m_p=1.293\,{\rm MeV}$ and $f_i=\{1+\exp[(E_i-\mu_i)/T]\}^{-1}$ is the Fermi-Dirac distribution. $C_V\simeq 1$ and $a=|C_A/C_V|\simeq 1.26$. 
Note here we consider} only the electron/positron processes (i.e., $n+e^+\to p+\bar\nu_e$ and $p+e^-\to n+\nu_e$) while neglecting the effect of the neutrino flux due to the lack of neutrino flux calculations; this issue can be addressed in separate studies where neutrino transport is considered.

The net production rates of nucleons in the gas phase are then given by
\begin{equation}
    \frac{dN_{n,p}}{dt} = \left[\frac{dN_{n,p}}{dt}\right]_{A\leftrightarrow (A-1) + n/p}+ \left[\frac{dN_{n,p}}{dt}\right]_{n\leftrightarrow p}\,,\label{eq: total dN}
\end{equation}
while the loss rate of baryon number is  
\begin{equation}\label{eq: dA}
    \frac{dA}{dt} = -\frac{dN_n}{dt}-\frac{dN_p}{dt}\,.
\end{equation}

Eqs.(\ref{eq: internal energy}, \ref{eq: total dN}, \ref{eq: dA}) constitute a system of differential equations. Once the initial values (i.e., $A_0, N_{n0}, N_{p0}, T_{s0}$), along with the trajectories of gas temperature $T(t)$ and total density $\rho(t)$, are provided, these equations can be evolved to yield detailed information about the composition and nugget temperature. Notably, the initial values can be derived from the equilibrium conditions, given the initial temperature $T_0=T(t=0)$ and the binding energy $\Delta E$.

In the above, we focus on the evolution of QNG. However, if $T_0$ is sufficiently high such that all quark nuggets evaporate into nucleons, we must instead begin with nucleon gas. In that case, only Eq.(\ref{eq: np conversion}) needs to be evolved.

\begin{figure}
 \centering
  \includegraphics[width=3in]{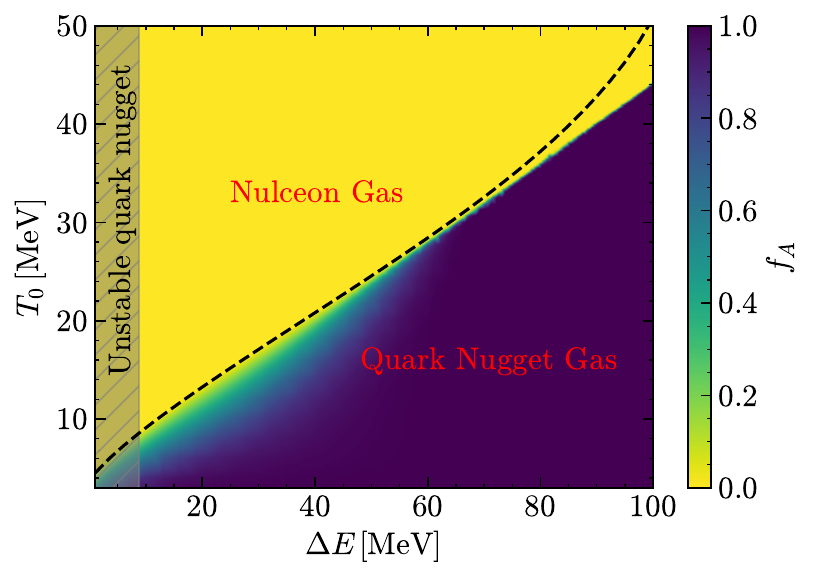} \hspace{0.in}%
    \caption{Final nugget fraction (when the gas temperature reaches $T=1\,{\rm MeV}$) for different binding energies $\Delta E$ and initial temperatures $T_0$. 
    Dashed line represents the critical boundary calculated using the equilibrium condition [Eq.(\ref{eq:critical line})]. The grey shaded region represents the regime $\Delta E < 8.8\,{\rm MeV}$, where quark matter is less stable than the $^{56}{\rm Fe}$ nucleus.}  
\label{fig:nugget fraction}
\end{figure}

{\it Application to homologous expansion of ejecta:} After a few milliseconds following the BQS merger, the ejecta could be considered to enter a homologous expansion stage, i.e., each part of the ejecta expands at a velocity proportional to its distance from the center, $r=v_{\rm ej} (t_h+t_0)$ with $v_{\rm ej}$ the velocity of the fluid element and $t_h(>0)$ the homologous expansion time, and $t_0$ is a constant that defines the $t_h=0$ point of homologous expansion. We then write the ejecta density and gas temperature as 
\begin{align}
    \rho(t_h) &= \frac{M_{\rm ej}}{(4\pi/3)v_{\rm ej}^3(t_h+t_0)^3},\label{eq:rho prof}\\
    T(t_h) &= T_0 \left(1+\frac{t_h}{t_0}\right)^{-3(\gamma-1)},\label{eq:temp prof}
\end{align}
where $\gamma$ is the adiabatic index~\citep{Endnotes}, $M_{\rm ej}$ and $v_{\rm ej}$ are the ejecta mass and velocity, respectively. 
In the following sample calculation, we set the initial density to $\rho_0 =\rho(0)=M_{\rm ej}/(\frac{4\pi}{3}v_{\rm ej}^3t_0^3) = 5\times 10^{12}\,{\rm g/cm^3}$, assuming that from this point the ejecta enters the homologous expansion stage.

Given the initial temperature $T_0$, we can evolve Eqs.(\ref{eq:rho prof}, \ref{eq:temp prof}) together with Eqs.(\ref{eq: internal energy}, \ref{eq: total dN}, \ref{eq: dA}). By doing so, we obtain the \textit{nugget fraction} $f_A=An_A/(An_A+n_n+n_p)$, representing the amount of matter contained in the nugget phase, and the \textit{proton fraction} $Y_p=n_p/(n_n+n_p)$, representing the proportion of protons in the gas phase.
Since nucleosynthesis (which is beyond the scope of this paper) starts to play a significant role around  $\sim 1\,\rm{MeV}$, and the evaporation efficiency becomes sufficiently low by that point, we stop evolving the system once the gas temperature reaches $1\,{\rm MeV}$. 

\begin{figure}
 \centering
  \includegraphics[width=3in]{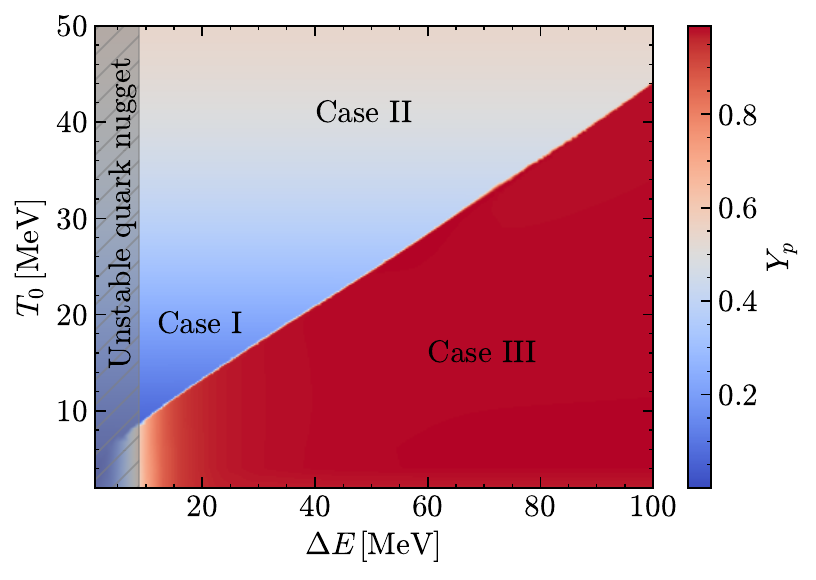} \hspace{0.in}%
    \caption{Same as Fig.~\ref{fig:nugget fraction}, but for the proton fraction in the gas phase.}  
\label{fig:proton fraction}
\end{figure}

In Fig.~\ref{fig:nugget fraction}, we show the ``final" nugget fraction when the gas temperature reaches where $T=1\,\rm{MeV}$, for different values of $T_0$ and $\Delta E$, while fixing $M_{\rm ej}=0.01\,M_\odot$, $v_{\rm ej}=0.1c$, $\gamma = 1/3$, $A_0 = A(t_h=0)=10^{30}$, and $E_c=6\,{\rm MeV}$. 
We see that the $T_0-\Delta E$ space is divided into two distinct regions: in the high $\Delta E$ and low $T_0$ region, $f_A > 0$, indicating that nuggets have not fully evaporated; in the low $\Delta E$ and high $T_0$ region, $f_A = 0$, which means that all nuggets have evaporated into nucleons.
This is consistent with our expectation: the higher the binding energy and the lower the temperature, the more quark nuggets remain un-evaporated.
The dashed line in Fig.~\ref{fig:nugget fraction} represents the boundary between QNG and NG as calculated from the equilibrium condition [Eq.(\ref{eq:critical line})]. However, when considering the non-equilibrium effects, the actual boundary lies slightly below the dashed line, as shown in Fig.~\ref{fig:nugget fraction}.

\begin{figure}
  \includegraphics[width=2.8in]{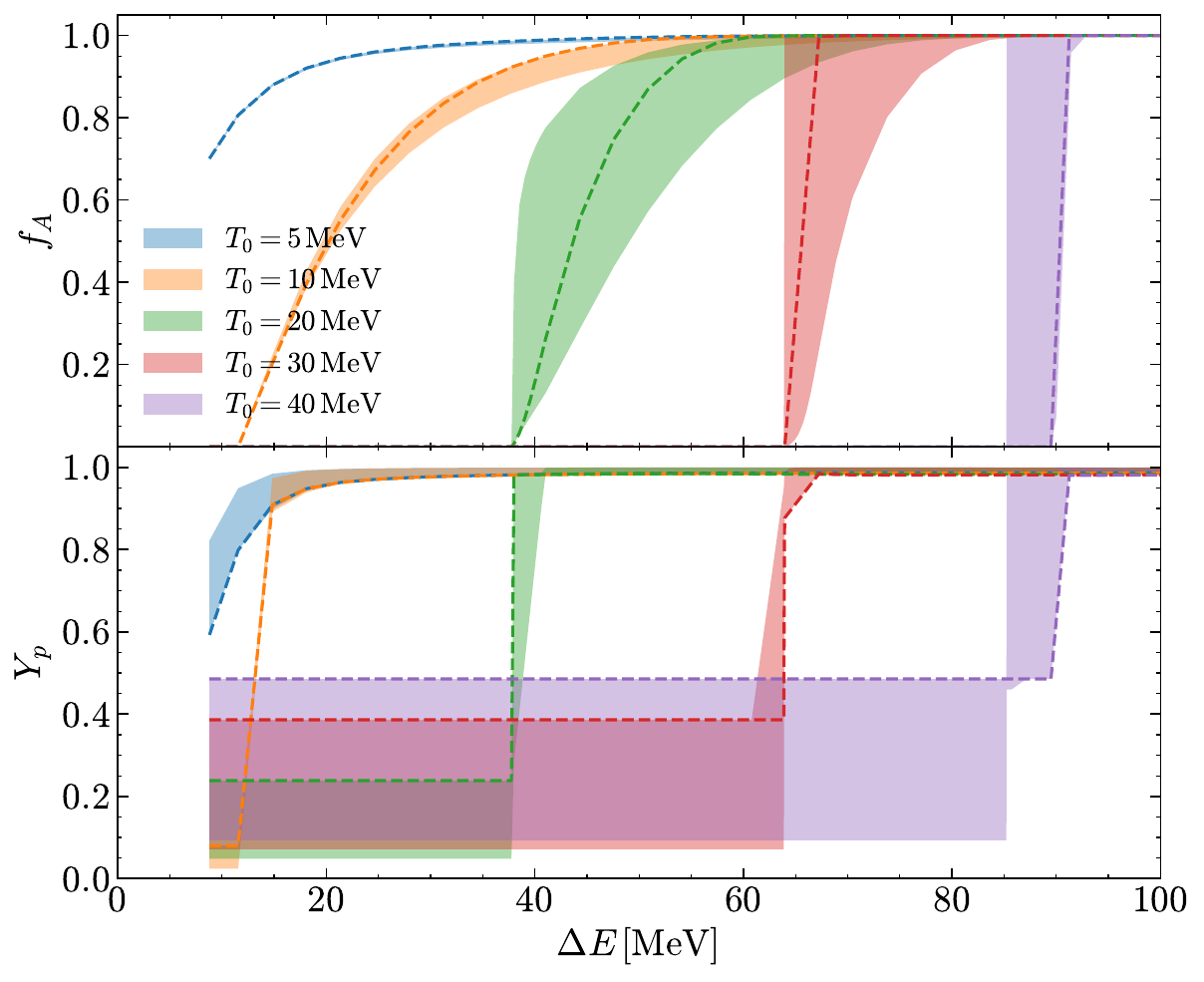} \hspace{0.in}%
    \caption{Final nugget fraction ($f_A$) and proton fraction ($Y_p$) as functions of binding energy $\Delta E$, for different initial temperatures. The dashed lines represent the results obtained from the parameter set used in Fig.~\ref{fig:nugget fraction} and Fig.~\ref{fig:proton fraction}. The solid bands represent the results when the parameters are varied over wide ranges (see text).}  
\label{fig:1d fractions}
\end{figure}

Fig.~\ref{fig:proton fraction} shows the corresponding proton fraction in the gas phase of the ejecta. 
We see that for small $\Delta E$,  $Y_p$ resembles that of a typical nucleon gas from BNS merger. In particular, in the lower-left region of Fig.~\ref{fig:proton fraction}, where $T_0$ is relatively low and $\Delta E$ is small, the gas is neutron-rich, as expected. 
However, $Y_p$ can be extremely high, approaching nearly 1, as $\Delta E$ increases.
The proton-rich nature is due to the fact that neutrons are more easily reabsorbed into the nugget, while protons face a greater obstacle in reabsorption because of the Coulomb barrier. 
Note that although the gas is extremely proton-rich in this case, since most mass is contained in the quark nuggets, the overall free proton fraction [$\propto (1-f_A)Y_p$] in the ejecta is very low.

Fig.~\ref{fig:1d fractions} illustrates the range of variation in $f_A$ and $Y_p$ as a function of $\Delta E$, for the parameter ranges $M_{\rm ej}\in [0.005, 0.02]\,M_\odot $, $v_{\rm ej} \in [0.1, 0.3]c$,  $\gamma \in [4/3, 5/3] $, and $E_c \in [6, 12]\,{\rm MeV}$.
We see that the variations of these parameters do not significantly alter the transition point from NG to QNG, they result in only moderate changes in the numerical values of $f_A$ or $Y_p$.

{\it Discussion and concluding remarks:} Whether BQS or QS-BH mergers can produce r-process elements and kilonova signals depends on whether the quark nuggets in the merger ejecta can effectively evaporate into nucleons. In this letter, we demonstrate that in the dense environment of BQS or QS-BH merger ejecta, evaporation is significantly suppressed due to saturation, and as long as the quark matter binding energy $\Delta E$ is sufficiently large, the evaporation efficiency remains very low for a wide range of parameter space, such that most of the mass in the ejecta is in the form of quark nuggets. 

\begin{figure}
  \includegraphics[width=2.8in]{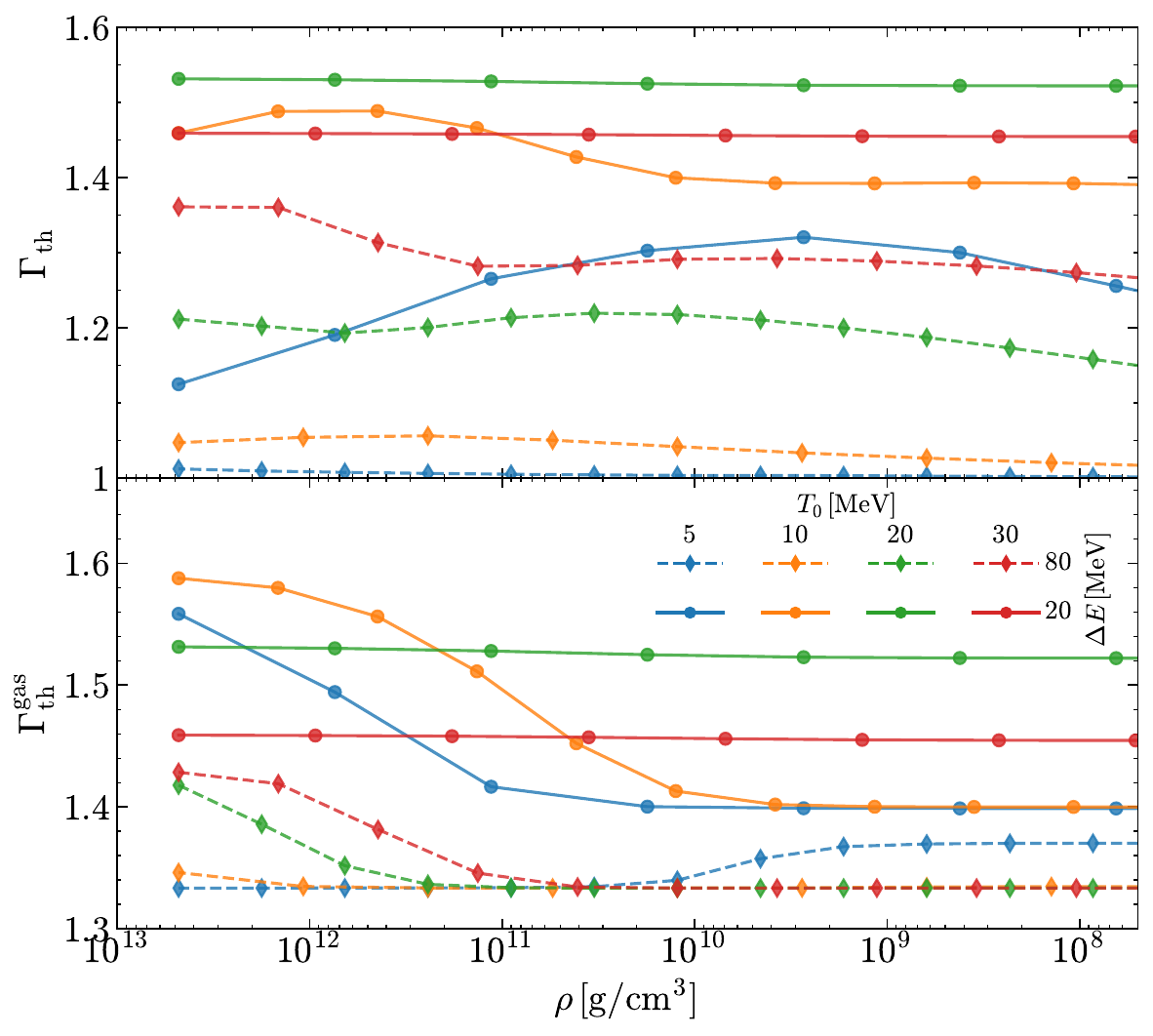} \hspace{0.in}%
    \caption{Thermal adiabatic index of the decompressed quark matter as a function of density. The upper panel shows $\Gamma_{\rm th} =P_{\rm th}/u_{\rm th}+1$, where $u_{\rm th}$ includes the internal thermal energy of the nuggets [see Eqs.(\ref{eq: pressure}--\ref{eq: e density})].
    The lower panel shows $\Gamma_{\rm th}^{\rm gas} =P_{\rm th}/u_{\rm th}^{\rm gas}+1$, where $u_{\rm th}^{\rm gas}=u_{\rm th}-u_A$ does not include the internal thermal energy of the nuggets.}  

\label{fig:thermal index}
\end{figure}

Our calculation of the ``final'' nugget fraction and proton fraction adopts a simple homologous expansion model for the ejecta evolution.
Nevertheless, we can draw several general conclusions.
There are three possible outcomes for the BQS merger ejecta (see in Fig.~\ref{fig:proton fraction}):
\begin{itemize}
    \item Case I: a low  $Y_p$ ($\lesssim 0.2$) nucleon gas,
    \item Case II: a medium $Y_p $ ($\sim 0.2-0.6$) nucleon gas,
    \item Case III: an extremely proton-rich ($Y_p \sim 1$) quark nugget gas.
\end{itemize}
In Case I, similar to BNS mergers, the neutron-rich ejecta can undergo r-process nucleosynthesis and produce a ``red'' kilonova.
In Case II, the ejecta may produce a ``blue'' kilonova.
In Case III, due to the small gas fraction ($1-f_A<10^{-2}$) and the extreme proton richness of the ejecta, no r-process nucleosynthesis and kilonova are expected.

To determine whether any of the above outcomes are produced in BQS or QS-BH mergers requires reliable knowledge of $\Delta E$ and more realistic temperature and density evolution of the ejecta, the latter must be provided by numerical simulations.
As a first step, we have applied the EoS formalism developed in this work to an existing BQS general relativistic hydrodynamical simulation from Ref.~\citep{2021PhRvD.104h3004Z}, and the results have confirmed the three cases discussed above (see Supplemental Material~\citep{SM}). 
However, we note previous simulations of BQS mergers (e.g., Refs.~\citep{2009PhRvL.103a1101B,2010PhRvD..81b4012B,2021PhRvD.104h3004Z,2022PhRvD.106j3030Z,2024arXiv240711143G}) did not treat the decompressed quark matter in a consistent way. 
In general, for QNG, adiabatic index, $\Gamma_{\rm th}=P_{\rm th}/u_{\rm th}+1$ is not constant, but evolves in time as the ejecta expands.
In Fig.~\ref{fig:thermal index}, we show the variation of $\Gamma_{\rm th}$ with density for our homologous expansion ejecta model. For future simulations of BQS or QS-BH mergers, a self-consistent treatment of $\Gamma_{\rm th}$ will be necessary to achieve reliable results.

Whether QSs exist in nature is unknown; it depends on the
quark matter binding energy $\Delta E$ (the energy required to release neutron from the bulk quark matter).
Theoretical calculations of the binding energy, through various approaches such as the bag model~\citep{2018PhRvD..97h3015Z,2021ApJ...917L..22M}, the Nambu–Jona–Lasinio model~\citep{2022PhRvD.105l3004Y}, the quark-meson model~\citep{2018PhRvL.120v2001H}, and others, remain highly uncertain. This uncertainty primarily arises from the poorly constrained QCD vacuum energy (the so-called ``bag parameter'', {$B$}), due to its nonperturbative nature. 
{For example, in the simplest bag model—neglecting the strange quark mass and gluon-mediated interactions—the binding energy is approximately related to $B$ via $\Delta E=m_n-(108\pi^2B)^{1/4}$. 
The ``standard value'' of $B$ is $(145\,{\rm MeV})^4$ from reproducing the mass spectrum of light hadrons and heavy mesons~\citep{1975PhRvD..12.2060D}, leading to $\Delta E=111\,{\rm MeV}$. 
In contrast, a relatively larger value of $B=(212\,{\rm MeV})^4$ is estimated from lattice calculations at zero chemical potential~\citep{1996NuPhA.606..320B}, which corresponds to $\Delta E=-272\,{\rm MeV}$, indicating that quark matter would be unstable. 
In Ref.~\citep{2021ApJ...917L..22M}, using tidal deformability constraint from GW170817 and assuming a BQS merger, an upper limit on $\Delta E$ was obtained within the framework of the bag model: $\Delta E \leq105\,{\rm MeV}$ (95\% confidence level).  
We refer to Ref.~\citep{2025arXiv250220241B} for recent theoretical progress on the bag parameter.}

In this Letter, our calculations show that for relatively high $\Delta E$ (e.g. $\gtrsim 50$~MeV), BQS or QS-BH mergers cannot produce neutron-rich ejecta in the gas phase, therefore cannot produce r-process elements and kilonova. 
In this scenario, AT2017gfo is unlikely to originate from a BQS merger. 
Alternatively, if it did, $\Delta E$ can be tightly constrained (e.g., $\Delta E\lesssim20\,{\rm MeV}$). 
Future NS-BH mergers observation may impose even tighter constraints on $\Delta E$, as temperatures in QS-BH mergers are generally lower. 
{On the other hand, the \textit{non-detection} of kilonovae in association with sufficiently nearby ``neutron star'' mergers could serve as potential evidence for the existence of QSs. However, robust conclusions require first ruling out alternative explanations within the NS framework---e.g., signals falling below detection thresholds due to ejecta mass, velocity, or opacity~\citep{2017Natur.551...80K}, which may be influenced by the nuclear EOS~\citep{2021ApJ...906...98N}, magnetic field effects~\citep{2024MNRAS.527.2240D}, nuclear reaction rates~\citep{2021ApJ...906...94Z,2023ApJ...944..144L}; the absence of tidal disruption in NS-BH mergers~\citep{2021ApJ...923L...2F}; contamination from bright optical afterglows or background emission from active galactic nucleus accretion disks~\citep{2022ApJ...938..147Z}. 
Disentangling these possibilities will be crucial in using kilonova observations--or their absence--to test the QS hypothesis.}

\acknowledgements{Acknowledgements: ZM would like to thank Ang Li for her valuable comments and Chen Zhang for helpful discussion regarding the binding energy. ZM is supported by the Postdoctoral Innovation Talent Support Program of CPSF (No. BX20240223) and the China Postdoctoral Science Foundation funded project (No. 2024M761948). ZZ is supported by the Postdoctoral Innovation Talent Support Program of CPSF (No. BX20220207), the National Natural Science Foundation of China (No. 12273028) and the China Postdoctoral Science Foundation funded project (No. 2022M712086).}

\input{SM}
\end{document}

%% file: SM.tex
\renewcommand{\thefigure}{S\arabic{figure}}
\renewcommand{\theequation}{S\arabic{equation}}



\section{Coulomb barrier at nugget surface and proton penetration factor}
Electrons are distributed more diffusely around the quark nugget compared to quarks, as they are bound by the electromagnetic force rather than the strong force. As a result, the nugget carries a net positive charge within its surface ($r=R_s$), with a surrounding electron ``halo'' outside. This creates a Coulomb barrier that a proton must overcome to penetrate the nugget. According to Refs.~[51, 53], the electrostatic potential outside the nugget is given by
\begin{equation}
    V(r) = \frac{3V_q}{\sqrt{6\alpha/\pi}V_q(r-R_s)+4}, \quad r>R_s.
\end{equation}
Here $V_q^3/(3\pi^2)$ is the quark charge density and can be estimated as  
$V_q\simeq m_s^2/4\mu_q\approx 8\,{\rm MeV}$ for $m_s=100\,{\rm MeV}$ and $\alpha=1/137$ is the fine-structure constant. 

We can calculate the proton penetration factor using the electrostatic potential $V(r)$ obtained above. 
The penetration factor is given by (see e.g., Ref.~[54])
\begin{equation}
    P = \frac{|\chi(R_s)|^2}{|\chi(\infty)|^2},
\end{equation}
where $\chi(r)$ is the solution of the radial wave equation
\begin{equation}
    \left[-\frac{1}{2m_p}\frac{d^2}{dr^2}+V(r)-E\right]\chi(r) = 0.
\end{equation}
We neglect the angular momentum because we work in the regime where $l(l+1)/2m_pr^2\ll V(r)$. 
Using the WKB approximation we solve the equation and obtain 

\begin{align}
    P(E>E_c) &= \frac{4kk^\prime}{(k+k^\prime)^2},\\
    P(E<E_c) &= \exp\left\{-2\sqrt{2m_p}\int_{R_s}^{R_0} \sqrt{V(r)-E} dr\right\}
\end{align}
where $k=\sqrt{2m_p(E-E_c)}$ and $k^\prime=\sqrt{2m_pE}$, with $E_c = V(R_s)=3V_q/4$ the height of Coulomb barrier at nugget surface. $R_0$ represents the turning point defined by $V(R_0)=E$.
In this work, since the temperature of interest ensures $E\gtrsim \mathcal{O}(1)\,{\rm MeV} \sim E_c$, we can carry out the approximation 
\begin{equation}
\begin{split}
      \langle \sigma_pv_p\rangle & \approx \langle P(E>E_c)\sigma_n v_p\rangle\\
      &=\sigma_n \int P(E>E_c)\sqrt{\frac{2E}{m_p}}\frac{2}{\sqrt{\pi}}\frac{E}{T} \exp\left(-\frac{E}{T}\right) \frac{dE}{(TE)^{1/2}} \\  
      &= I_p\sigma_n \sqrt{\frac{8T}{\pi m_p}} ,
\end{split}
\end{equation}
where 
\begin{equation}
    I_p=\int_\eta^\infty \frac{4\sqrt{1-\eta/\xi}}{(1+\sqrt{1-\eta/\xi})^2}\xi e^{-\xi}d\xi,
\end{equation}
with $\xi= E/T$ and $\eta = E_c/T$.
\begin{figure}
 \centering
  \includegraphics[width=3.in]{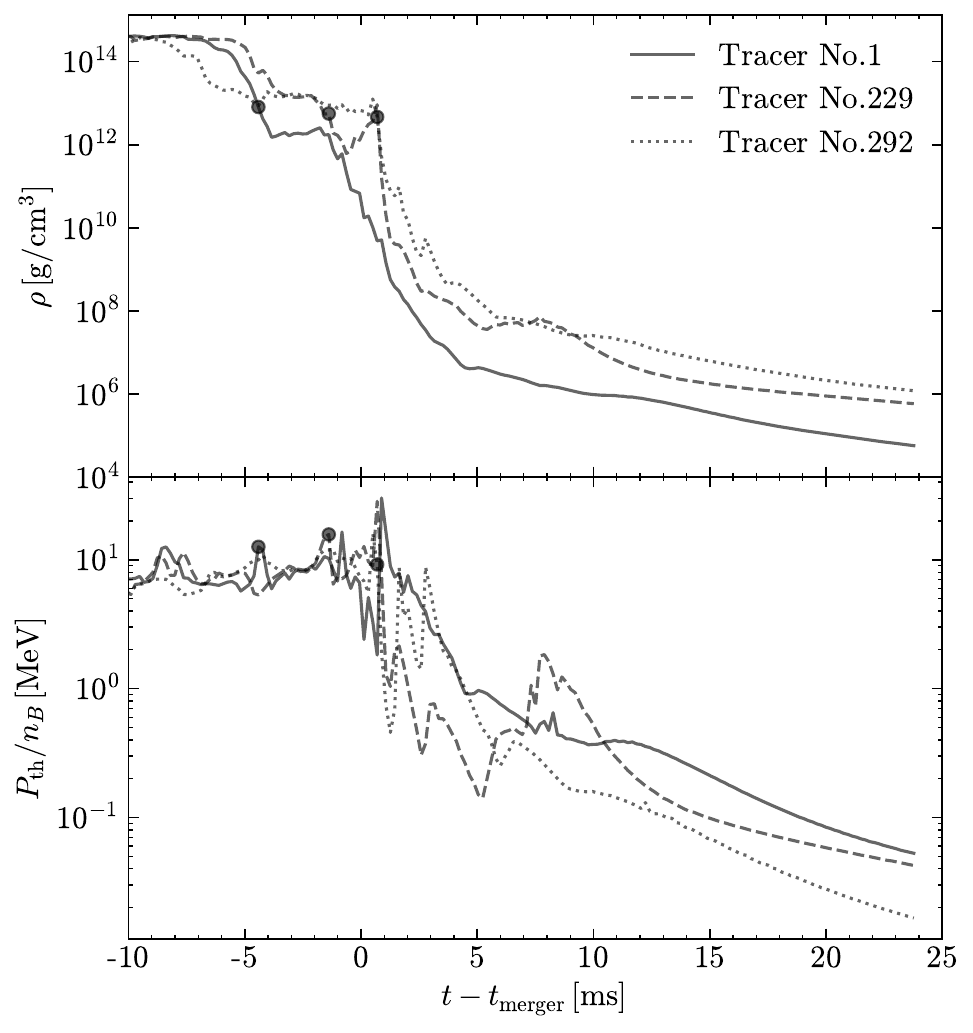} 
    \caption{Density and pressure evolution tracks for three representative tracers. The black points represents the starting points chosen in our calculations.}  \label{fig: tracers}
\end{figure}

\section{Application to binary quark-star merger simulations}

In this section, we use the recent BQS simulation results from Ref.~[42] as the input to analyze the properties of the ejecta. The input data from the simulations include the density ($\rho$) and pressure ($P_{\rm th}$) evolution tracks for different tracers. 
There are a total of 1030 tracers, and three representative tracers are shown in Fig.~\ref{fig: tracers}.

\begin{figure*}[hbt]
 \centering
  \includegraphics[width=4.in]{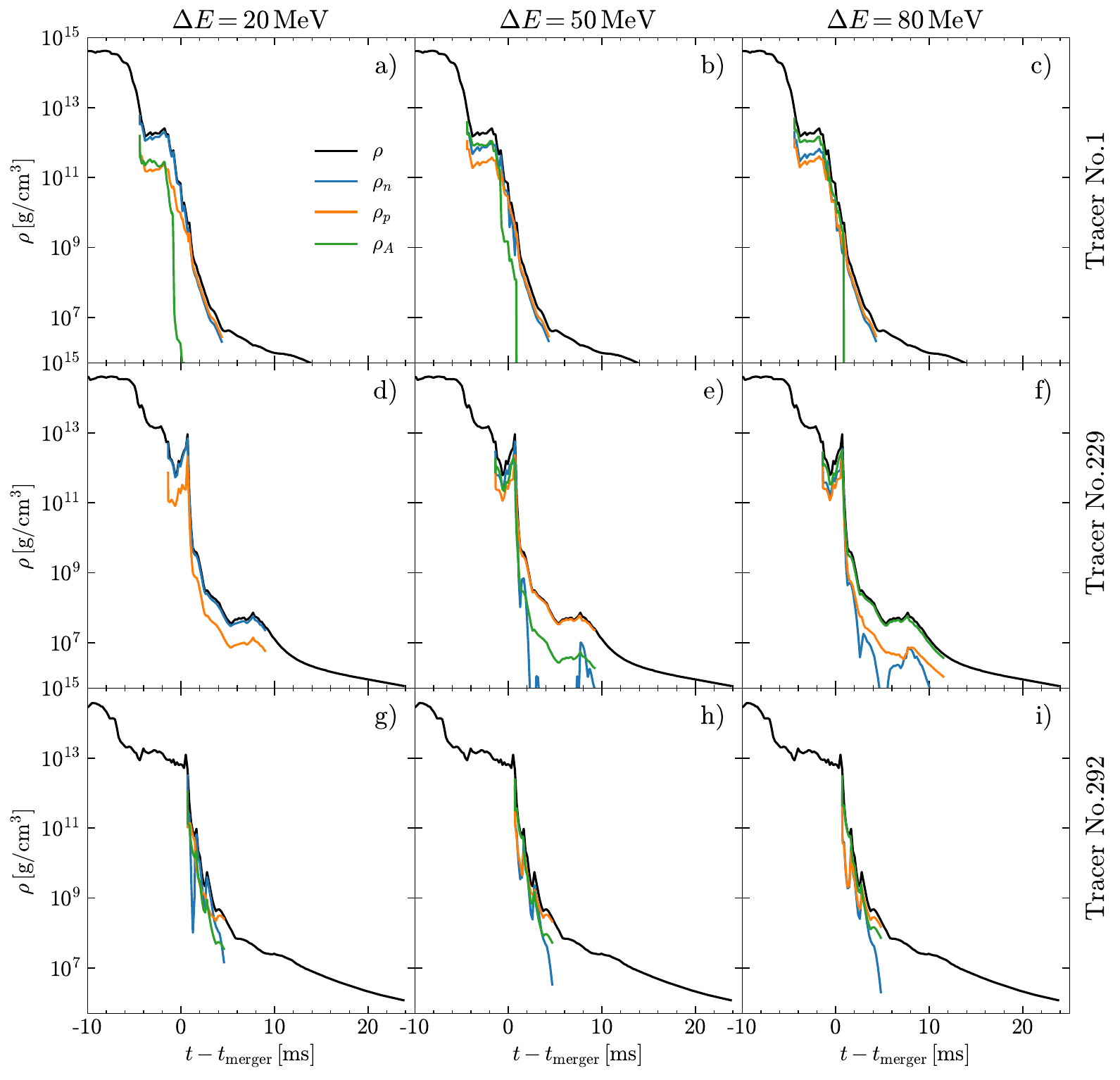} 
    \caption{Densities of neutrons ($\rho_n$), protons ($\rho_p$), quark nuggets ($\rho_A$), as well as the total density ($\rho$) for three representative tracers as a function of time. The results are obtained by using the density and pressure evolution tracks from a BQS simulation in Ref.~[42], and assuming different quark matter binding energy $\Delta E$.} \label{fig:density profile}
\end{figure*}

\begin{figure*}[]
 \centering
  \includegraphics[width=4.in]{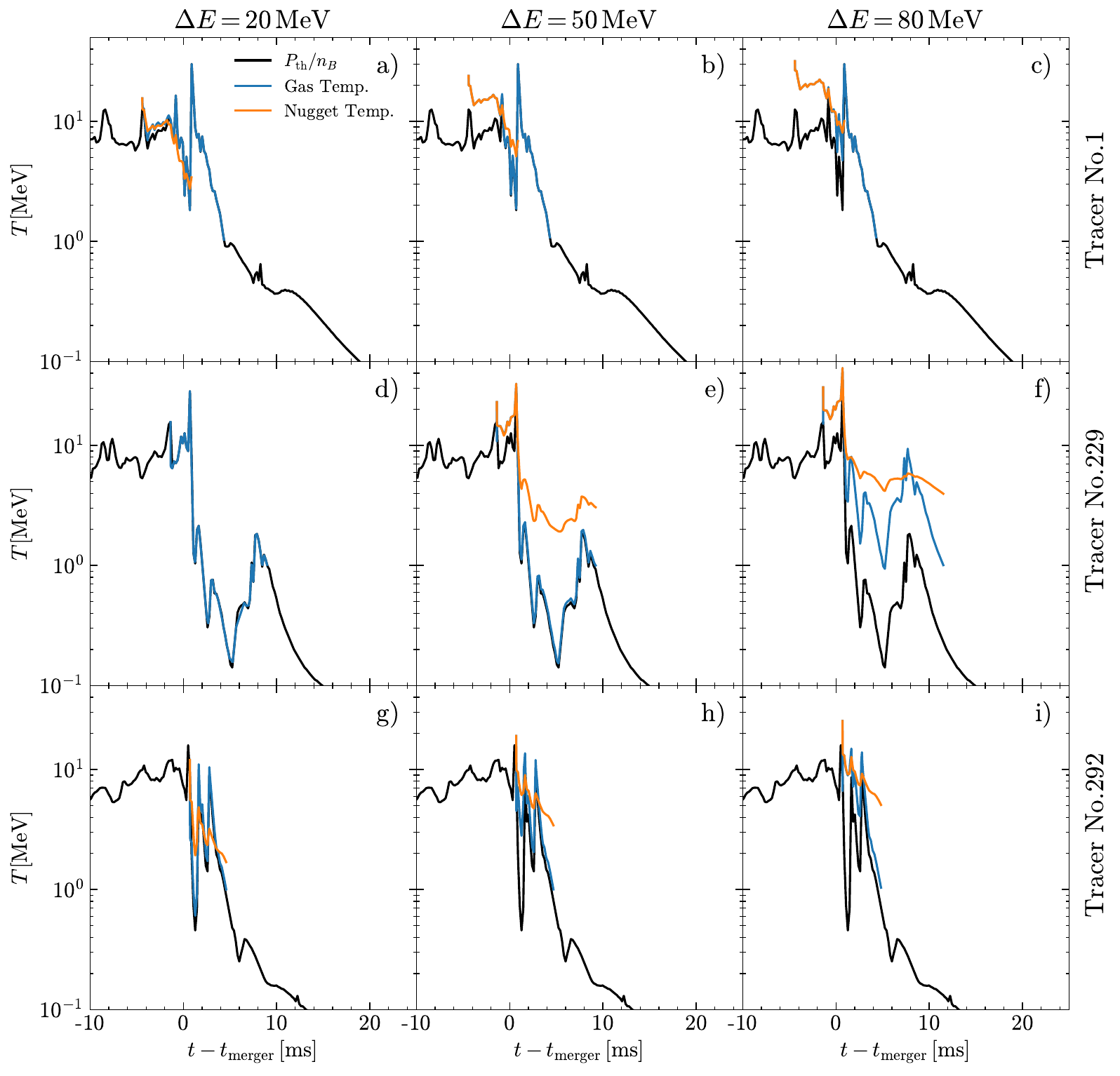} 
    \caption{Same as Fig.~\ref{fig:density profile}, but for the gas temperature and the nugget internal temperature. }  \label{fig:temp profile}
\end{figure*}

To calculate the ejecta properties, we first select the starting points where the density is approximately $10^{12}\,{\rm g/cm^3}$; these are marked as black points in Fig.~\ref{fig: tracers}. Using the equilibrium EoS formalism, we obtain the initial values of ($N_n, N_p, T_s$), with $A = 10^{30}$. We then evolve each tracer system by using the formalism developed in the main text for QNG or NG, and stop the integration when the gas temperature reaches $1\,{\rm MeV}$. 
At each iteration step, the gas temperature $T$ is determined by $P_{\rm th} = (n_n + n_p)T$. 
We do not include $P_\gamma+P_{e^+e^-}$ when inferring gas temperature from pressure, as the simulation in Ref.~[42] considered only the fluid contribution to the pressure.
Fig.~\ref{fig:density profile} and Fig.~\ref{fig:temp profile} show the density and pressure evolution tracks for three representative tracers. 

\begin{figure*}
 \centering
  \includegraphics[width=3.5in]{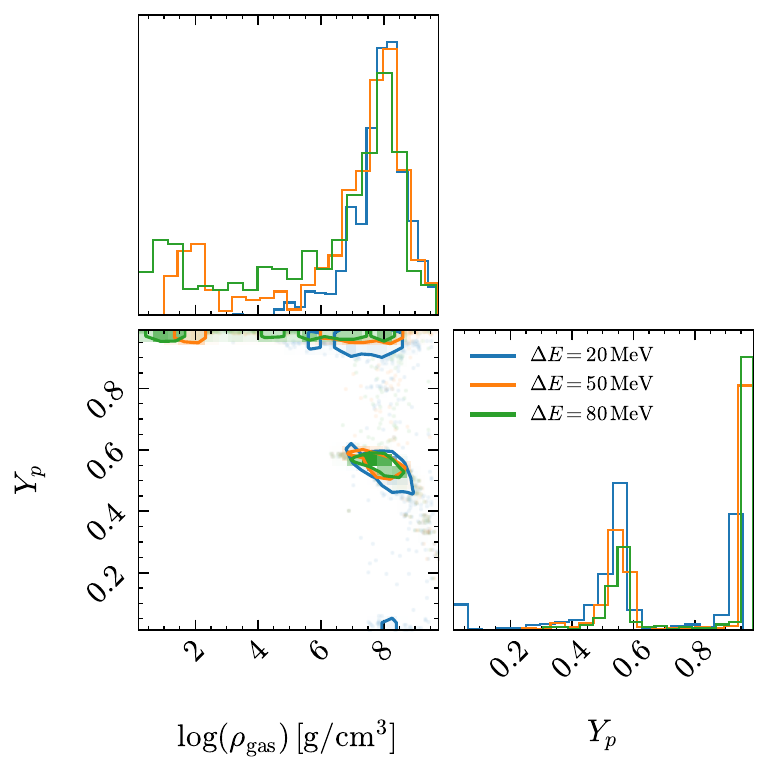} 
    \caption{Distributions of the final gas density ($\rho_{\rm gas}$) and proton fraction ($Y_p$) when the gas temperature reaches $T=1\,{\rm MeV}$. The results are obtained by using the density and pressure evolution tracks from a BQS simulation in Ref.~[42], and assuming different quark matter binding energy $\Delta E$.}  \label{fig:distribution}
\end{figure*}

In Fig.~\ref{fig:distribution}, we present the distributions of the gas density $\rho_{\rm gas}=\rho_n+\rho_p$ and the proton fraction $Y_p$, which are important inputs for subsequent nucleosynthesis calculations. We see that the $\rho_{\rm gas}-Y_p$ distribution is primarily concentrated in three regions, which correspond to the three possible outcomes, Case I--III, as discussed in the main text. This agreement supports the discussions and conclusions presented in the main text of the paper.

%% file: main.bbl
\begin{thebibliography}{99}
\bibitem[]{Endnotes} Here, $\gamma$ is not exactly the same as the thermal index  $\Gamma_{\rm th}$ in the EoS of the decompressed quark matter. We introduce it here with the aim of describing the expansion of the ejecta and treat it as a free parameter.
\end{thebibliography}
